\documentclass[11pt,twoside]{article}
\usepackage{asp2014}

\aspSuppressVolSlug
\resetcounters

\begin{document}

\title{Post-AGB evolution much faster than previously thought}
\author{K. Gesicki$^1$, A. A. Zijlstra$^2$ and M. M. Miller Bertolami$^3$
\affil{$^1$Centre for Astronomy, Faculty of Physics, Astronomy and
  Informatics, Nicolaus Copernicus University, Grudziadzka 5, PL-87-100 Torun,
  Poland; \email{kmgesicki@umk.pl}} 
\affil{$^2$Jodrell Bank Centre for Astrophysics, School of Physics
  \&\ Astronomy, University of Manchester, Oxford Road, Manchester M13\,9PL,
  UK; \email{a.zijlstra@manchester.ac.uk}}} 
\affil{$^3$Instituto de Astrof\'isica de La Plata, UNLP-CONICET, Paseo del Bosque s/n, 1900 La Plata, Argentina; \email{marcelo@mpa-garching.mpg.de}}

\paperauthor{K. Gesicki}{Krzysztof.Gesicki@astri.uni.torun.pl}{​0000-0003-4396-9527}{Centre for Astronomy, Faculty of Physics, Astronomy and Informatics}{Nicolaus Copernicus University}{Torun}{ }{PL-87-100}{Poland}
\paperauthor{A. A. Zijlstra}{a.zijlstra@manchester.ac.uk}{ }{Jodrell Bank Centre for Astrophysics, School of Physics \&\ Astronomy}{University of Manchester}{Manchester}{ }{M13 9PL}{UK}
\paperauthor{M. M. Miller Bertolami}{marcelo@mpa-garching.mpg.de}{ }{Instituto de Astrofísica de La Plata}{UNLP-CONICET}{La Plata}{ }{1900}{Argentina}

\begin{abstract}
For 32 central stars of PNe we present their parameters interpolated among the new evolutionary sequences. The derived stellar final masses are confined between 0.53 and 0.58 $M_\odot$ in good agreement with the peak in the white dwarf mass distribution. Consequently, the inferred star formation history of the Galactic bulge is well restricted between 3 and 11\,Gyr and is compatible with other published studies. The new evolutionary tracks proved a very good as a tool for analysis of late stages of stars life. The result provide a compelling confirmation of the accelerated post-AGB evolution.
\end{abstract}

\section{Introduction}

The main stellar evolutionary tracks for post-AGB evolution are now
over 20 years old. Recent results show that they predict the wrong
masses and ages for central stars of planetary nebulae (PNe).
\citet{GZHS2014} based on a detailed analysis of 31 Galactic bulge PNe
proposed that the post-AGB evolution should be accelerated by a factor
of 3. Coincidentally, during the last White Dwarf Workshop,
\citet{M3B2015} presented new models for the evolution of central
stars of PNe obtaining similarly shorter post-AGB timescales on purely
theoretical grounds. Here we re-analyse the data sample (recently
partially upgraded) however this time we interpolate among the new
evolutionary sequences aiming to verify their timescales.

\section{Methods and results}

The new evolutionary models were published by \citet{M3B2016}. They included all the recently available improvements to the TP-AGB evolution and are very adequate to the discussion of central stars of planetary nebulae (cspn). In Fig.\,\ref{agte} we show the post-AGB stages of evolution for six new tracks computed for solar metallicity. For comparison we show three old evolutionary sequences from \citet{B1995} -- their slower evolution is apparent.

\articlefigure[width=13.cm]{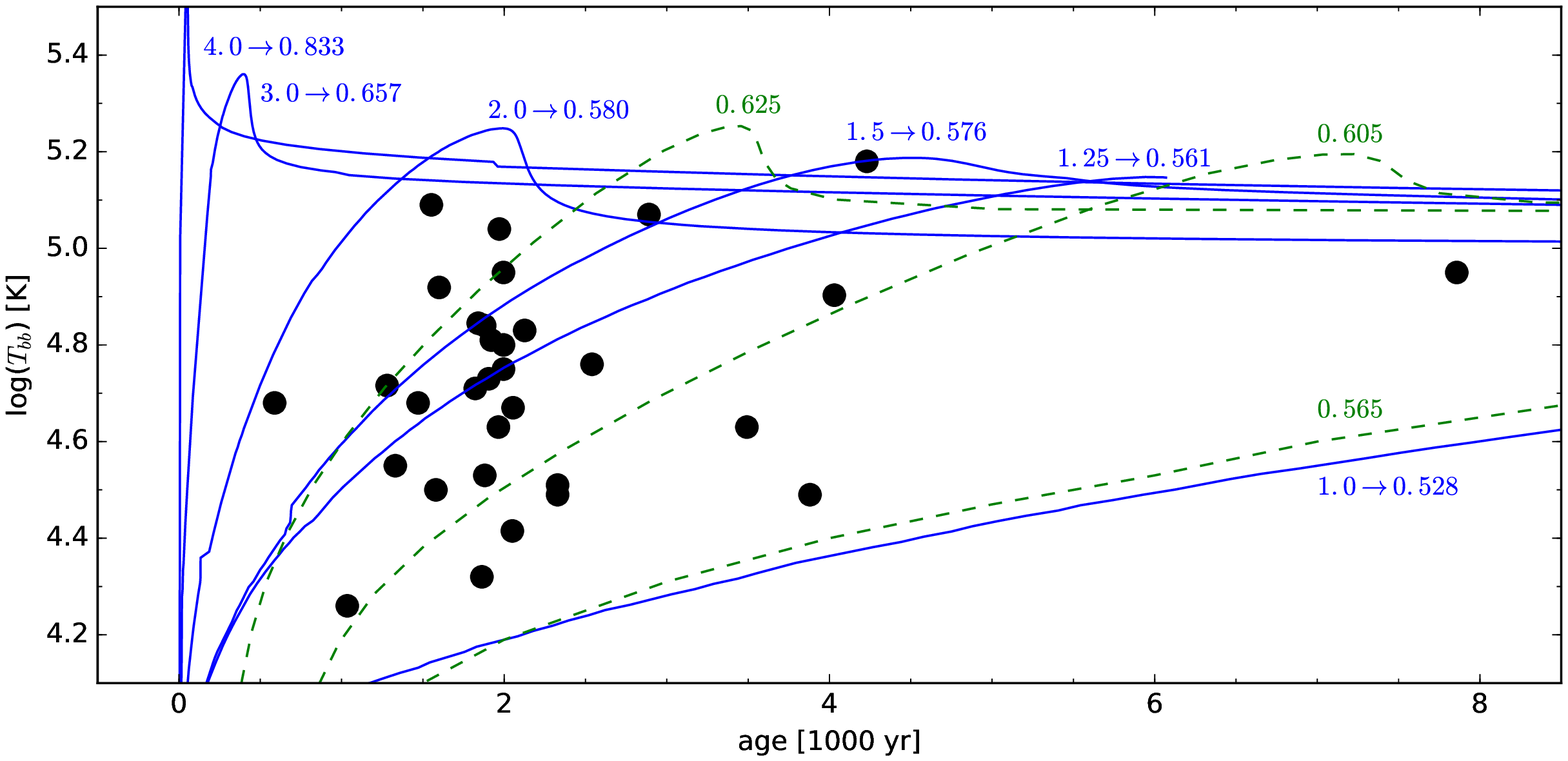}{agte}{Stellar temperature vs. post-AGB age for the derived data points and for different evolutionary calculations. The solid (blue) lines show the recent evolutionary calculations of \citet{M3B2016} for $Z=0.02$, the lines are labelled with their initial and final masses connected by an arrow. For a comparison the dashed (green) lines show three old evolutionary sequences of \citet{B1995} labelled with their final mass. }

The sample of 31 PNe \citep{GZHS2014} was recently partially upgraded after the careful 3D modelling published in \citet{GZM2016}: one object has been added and parameters of seven others were modified. The parameters derived from photoionization/kinematical reconstruction are: stellar black-body temperature and nebular kinematic age. The 32 data points are shown in Fig.\,\ref{agte}. Relating these points with grids of evolutionary models allows for interpolation of stellar masses (both initial and final) and total ages counted from the ZAMS; results are presented in histograms in Fig.\,\ref{histo}.

\articlefigure[width=13.cm]{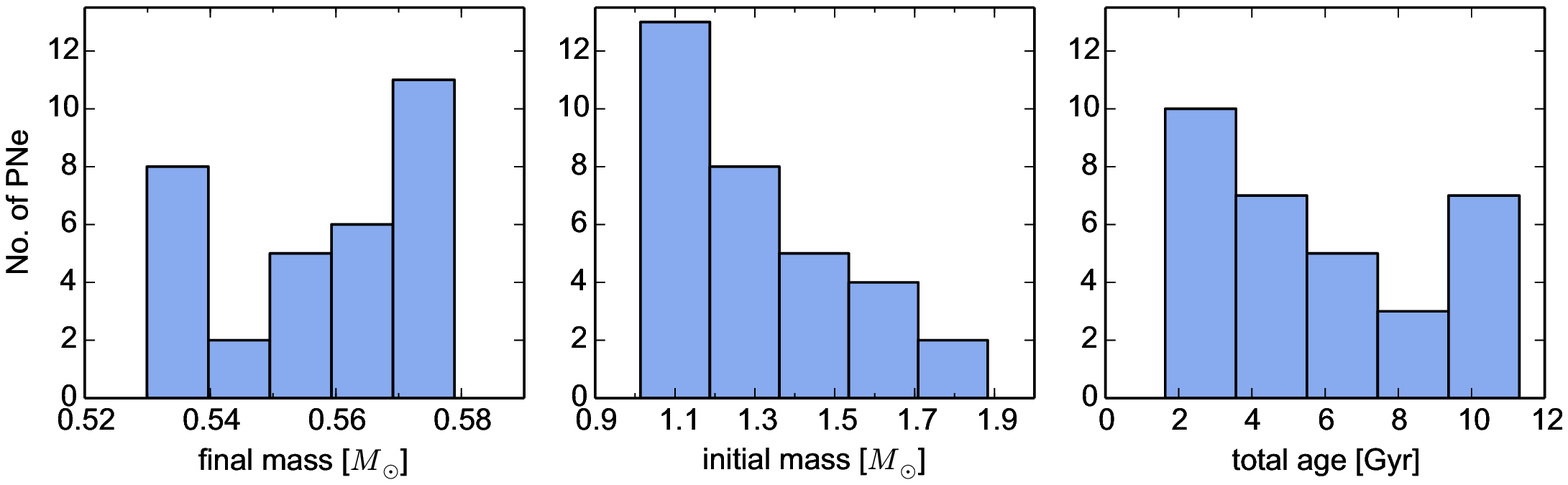}{histo}{Histograms of cspn final masses, initial masses and total ages. These parameters are derived from the PNe modelling data and the new evolutionary tracks.}

\section{Discussion}

The 32 data points in Fig.\,\ref{agte} represent the most carefully analysed Milky Way bulge planetary nebulae. Because of the limited number of objects, the histograms shown in Fig.\,\ref{histo} are composed of five bins only, equally separated between the extremes. 

\subsection{Final masses}

All derived cspn masses fall within an interval of 0.05 $M_\odot$ which very often is a single bin for WD mass histograms, see e.g. \citet{T2016}. Therefore the comparison with WD distributions is not perfect, nevertheless all our cspn masses agree with the main peak ($0.55-0.60\,M_\odot$) of the WD masses in the well-studied samples.

The analysed PNe data points cluster in Fig.\,\ref{agte} near post-AGB ages of 2\,kyr. The interpolation of \citet{GZ2007} between equally spaced old Bl\"ocker tracks resulted in a distribution sharply peaked at 0.61\,$M_\odot$ while the accelerated Bl\"ocker tracks  of \citet{GZHS2014} show the peak shifted to 0.58\,$M_\odot$. The new tracks of \citet{M3B2016} reveal that the area most populated by PNe is traversed by evolutionary tracks of a range of initial (and final) masses. In consequence the obvious peak no longer exists and the final mass histogram (Fig.\,\ref{histo} left panel) is broadened.

\subsection{Initial masses and total ages}

To derive the initial masses we adopted the theoretical Initial-Final Mass
Relation (IFMR) from the new tracks \citep{M3B2016}. This IFMR
displays a pronounced plateau in the range of masses of interest here,
with models of initial masses between 1.25 and 2.0\,$M_\odot$ all ending up
with masses between 0.56 and 0.58\,$M_\odot$. This feature is common to many
theoretical IFMRs.

About half of our sample falls between the 1.0 and 1.25\,$M_\odot$ tracks (ages above 5\,Gyr), while the other half is spread between the tracks corresponding to 1.25--2.0\,$M_\odot$ (ages 5--1.5\,Gyr respectively), gradually decreasing towards higher masses. Due to the plateau in the IFMR, the concentration of cspn with masses between 0.56 and 0.58\,$M_\odot$ (Fig.\,\ref{histo} left panel) is spread over a wide range of initial masses (Fig.\,\ref{histo} middle panel). One quarter of the sample has final masses $\sim0.535\,M_\odot$, which corresponds to initial masses around $\sim1.05\,M_\odot$ and ages of $\sim10\,$Gyr -- these cspn correspond to the peak at very old ages (Fig.\,\ref{histo} right panel).

\subsection{Ages and star-formation history}

The histogram of total ages was used to derive a star-formation history (SFH). The corrections for PNe visibility-time and for PNe birth-rate were applied exactly as in \citet{GZHS2014}. These corrections suppress the bins at young ages while the old ages bins are enhanced. The resulting SFH is shown in Fig.\,\ref{sfr} as a normalized histogram. Obviously there is some uncertainty related with those corrections and their details should be improved in a future.

\articlefigure[width=9.cm]{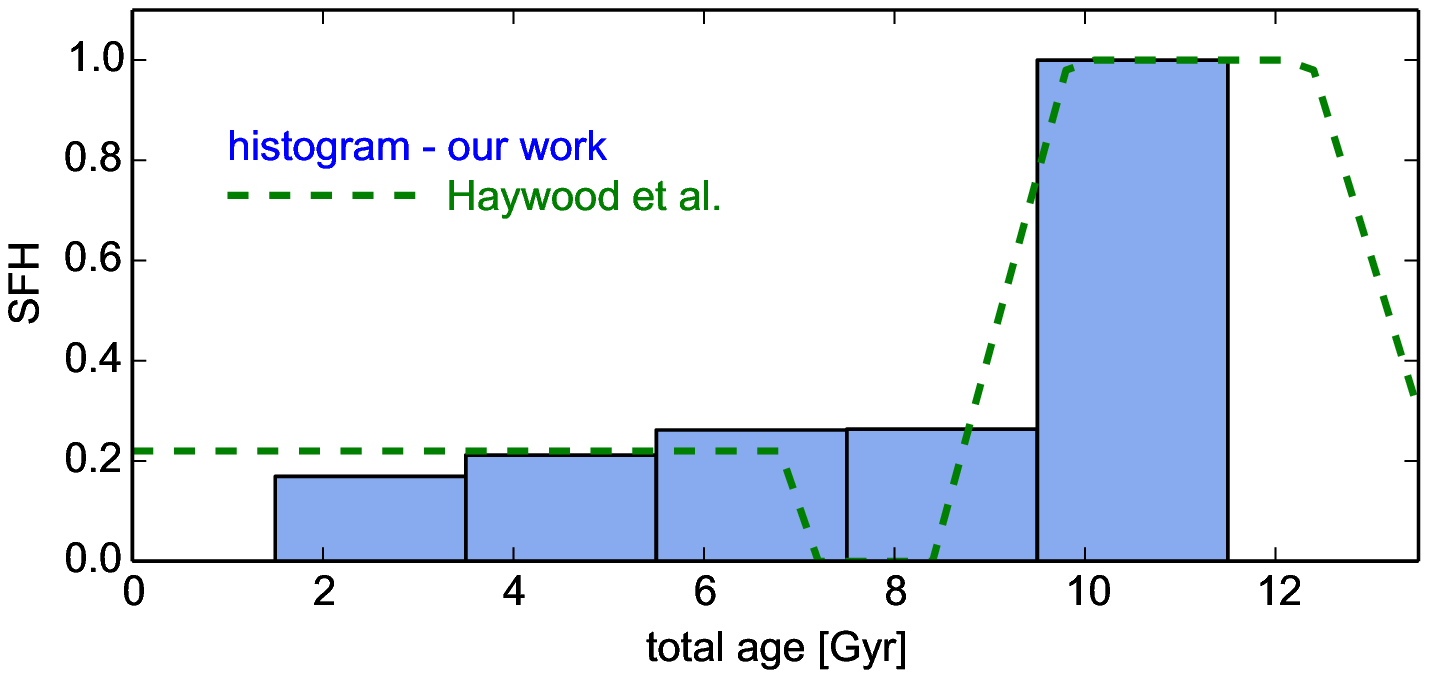}{sfr}{Normalized to unity star-formation-history derived from the histogram of the total ages corrected by PN visibility and birth-rate. The dashed green line shows, for comparison, the SFH derived by \citet{Ha2016}.}

There is in literature a general consensus that low metallicity bulge stars are older than $\sim10$\,Gyr while for super-solar metallicity there is no such consensus; different and often contradicting arguments are neatly summarized by \citet{N2016}. With the use of chemical evolution models \citet{Hb2016} showed that the SFH and the age-metallicity relation obtained for the Galactic inner disk \citep{Ha2016} provided also a best fitting color-magnitude diagram for the Galactic bulge. Their conclusion was that the metal-rich stars of the bulge must be significantly younger than the (still dominating) metal-poor population and this is an evidence for the bulge formed through the dynamical instabilities in the disk. Interestingly there is a remarkable similarity between their SFH and ours, both are compared in Fig.\,\ref{sfr}.

\section{Conclusions}

The most important outcome of our work are the much lower final masses, as compared with the old models, which is a robust result. The obtained final masses are in good correspondence with the most common white dwarf masses, $\sim0.55\,M_\odot$. Our results agree well with the previous suggestion \citep{GZHS2014} that the \citet{B1995} tracks were far too slow. The Galactic bulge SFH, derived from the total ages, points to a burst-like old population, plus some younger objects which is in agreement with the recent Galactic disk SFH and with the idea that the bulge originated from disk instabilities. These numerous accordances constitute a strong support for the short timescales of the new models of \citet{M3B2016}. The kinematical/photoionization reconstruction method elaborated recently by \citet{GZM2016} proved to be efficient.

\acknowledgements K.G. acknowledges financial support by Uniwersytet Mikolaja
Kopernika w Toruniu. M3B is partially supported by ANPCyT through grant PICT-2014-2708 and by a Return Fellowship from the Alexander von Humboldt Foundation.



\end{document}